\documentclass[reprint,superscriptaddress,amsmath,amssymb,aps,longbibliography,floatfix]{revtex4-2}
\usepackage{graphicx}
\usepackage[normalem]{ulem}
\usepackage[svgnames]{xcolor}
\usepackage{bm}
\usepackage{multirow}
\usepackage{float}
\restylefloat{table}
\usepackage{titlesec}
\usepackage[columnwise]{lineno}
\usepackage{titlesec}
\usepackage{gensymb}
\usepackage{textcomp}
\begin{document}
\title{T-square dependence of the electronic thermal resistivity in metallic strontium titanate}

\author{Shan Jiang}
\affiliation{Laboratoire de Physique et d'Étude des Matériaux\\ (ESPCI Paris - CNRS - Sorbonne Universit\'e), PSL University, 75005 Paris, France}

\author{Beno\^it Fauqu\'e}
\affiliation{JEIP, USR 3573 CNRS, Coll\`ege de France, PSL University, 75231 Paris Cedex 05, France}
\author{Kamran Behnia} 
\affiliation{Laboratoire de Physique et d'Étude des Matériaux\\ (ESPCI Paris - CNRS - Sorbonne Universit\'e), PSL University, 75005 Paris, France} 

\date{\today}
\begin{abstract}

The temperature dependence of the phase space for electron-electron (e-e) collisions leads to a T-square contribution to electrical resistivity of metals. Umklapp scattering are identified as the origin of momentum loss due to e-e scattering in dense metals. However, in dilute metals like lightly doped strontium titanate, the origin of T-square electrical resistivity  in absence of Umklapp events is yet to be pinned down.  Here, by separating electron and phonon contributions to heat transport, we extract the electronic thermal resistivity in niobium-doped strontium titanate and show that it also displays a T-square temperature dependence. Its amplitude correlates with the T-square electrical resistivity. The Wiedemann-Franz law strictly holds in the zero-temperature limit, but not at finite-temperature, because the  two T-square prefactors are different by a factor of $\approx 3$, like in other Fermi liquids. Recalling the case of $^3$He, we argue that T-square thermal resistivity does not require Umklapp events. The approximate recovery of the Wiedemann-Franz law in presence of disorder would account for a T-square electrical resistivity without Umklapp.
\end{abstract}
\maketitle

%Fermi liquids display quadratic temperature dependence resistivity at sufficiently low temperatures with two known mechanisms: multiple electron reservoirs and Umklapp scattering. While T-square resistivity found in dilute metal SrTiO$_3$ and Bi$_2$O$_2$Se in the absence of the two mechanisms attracts recent attention. 
Landau and  Pomeranchuk \cite{Landau1936}, and contemporaneously Baber \cite{Baber1937} postulated that electron-electron collisions cause a quadratic temperature dependence in electrical resistivity of metals. Subsequent experiments found that this is prominent in metals hosting strongly-correlated electrons (Such as UPt$_3$ \cite{Joynt2002} or strontium titanate \cite{Maeno1997}), but also those with a small carrier concentration (like bismuth \cite{Hartman1969} and graphite \cite{Uher1977}). In these cases, at sufficiently low temperature, resistivity, $\rho$, can be expressed as :

\begin{equation}
  \rho = \rho_0 + A T^2
  \label{T-square1}
\end{equation}

Here, $\rho_0$ is the residual resistivity, which  depends on disorder. The prefactor, $A$, on the other hand is intrinsic to each metal. The ubiquity of equation \ref{T-square1} across various families of Fermi liquids raises two questions: 1) What makes the exchange of momentum between two colliding electrons detrimental to the electrical conduction? 2) What sets the amplitude of $A$? 

The two identified answers to the first question are Umklapp and the Baber mechanism. An Umklapp event occurs when the momentum vector sum of the colliding electrons gets out of the Brillouin zone, leading to a loss of momentum equivalent to one reciprocal unit vector \cite{Yamada1986, Maebashi1998}. The Baber mechanism \cite{Baber1937} refers to the existence of two distinct electron reservoirs whose momentum exchange is a bottleneck in the path of momentum leak from the electron bath to the phonon bath. The second question was addressed first by Rice \cite{Rice1968} and then by Kadowaki and Woods \cite{Kadowaki1986} (See also  \cite{Tsujii2003,Hussey2005}) who argued that the amplitude of $A$ scales with the square of the $T$-linear specific heat, $\gamma^2$, because both depend on the density of states. 

The persistence of T-square electric resistivity in  metallic strontium titanate (STO) \cite{Okuda2001,Marel2011} to the extreme dilute limit \cite{Lin2015} raised new questions about these answers. The Fermi surface in this dilute metal is too small to allow Umklapp scattering and it consists of a single pocket in the extreme dilute limit \cite{collignon2019,Lin2013,Lin2014}. Thus, none of the two mechanisms can generate  T-square resistivity. Moreover, the proper scaling relation was found to be between $A$ and the Fermi energy, $A \propto E_{F}^{-2}$, instead of the standard Kadowaki-Woods scaling, $A \propto \gamma^2$, which fails in STO \cite{McCalla}.

Following this observation, transport properties of dilute metallic STO were studied up to temperatures well above room temperature \cite{Collignon2020} and the effective mass was found to increase with warming. Two theoretical studies \cite{Kumar2021, Nazaryan2021} showed that the temperature dependence of electrical resistivity in STO one can be explained with a scenario based on the scattering of electrons by two soft transverse optical (TO) phonons. This would account for the persistence of T-square resistivity above the degeneracy temperature. On the other hand, low-temperature T-square resistivity (well below  the minimum energy of TO phonons) remained a mystery. Another development was the discovery of  T-square resistivity in dilute metallic Bi$_2$O$_2$Se \cite{Wang2020}, another solid with a small Fermi surface, and without any soft phonon mode. The T-square prefactor was found to  scale with the Fermi energy. This demonstrated that STO is not an isolated case and called for an e-e scattering scenario in absence of Umklapp.

%(i.e., those having roughly one carrier per formula unit) or one over the square of Fermi energy $E_F$ ($A \propto E_{F}^{-2}$) for including dilute metals\cite{Behnia2022, Jaoui2021, Lin2015, Wang2020}. The participating electrons are confined near the Fermi energy, leading to $T$-square phase space for electron-electron collision.

The T-square \textit{thermal} resistivity of electrons in Fermi liquids is less known and even much less explored. Defining the electronic thermal resistivity as $WT = T/\kappa^{e}$, one expects:
\begin{equation}
  WT = (WT)_0 + B T^2
  \label{T-square2}
\end{equation}

Here, $(WT)_0$ is the residual thermal resistivity, expected to obey the Wiedemann Franz (WF) law: $L_0(WT)_0 = \rho_0$ with $L_0 = \frac{\pi^2}{3} \frac{k_B^2}{e^2}=2.44\times10^{-8} V^2K^{-2}$.  On the other hand,  $L_0B > A$, because, compared to energy flow, momentum flow is less affected by small-angle scattering \cite{ziman1972, Li-Maslov}. Experiments on various metals, including Ni \cite{White1967}, Al \cite{Garland1978}, W\cite{Wagner1971}, Sb \cite{Jaoui2021}, CeRhIn$_5$ \cite{Paglione2005}, WP$_2$ \cite{Jaoui2018}, UPt$_3$ \cite{Lussier1994}) have confirmed both these expectations.  

A quantitative connection between this physics and heat transport in the normal liquid $^3$He was recently highlighted \cite{Behnia2022}. In normal liquid $^3$He , thermal conductivity becomes proportional to the inverse of temperature \cite{greywall1984} at very low temperatures, which means that thermal resistivity $WT = T/\kappa$, is proportional to $T^2$. The evolution of this T-square  resistivity  with pressure follows the scaling  seen for $A$ and $B$ with $E_F$ in metals \cite{Behnia2022}. There is no Umklapp in normal liquid $^3$He and the Fermi surface is a single sphere. Thus, T-square thermal resistivity can occur in a Fermi liquid without Umklapp and the amplitude of $B$ is directly linked to its Landau parameters, which also set the Fermi temperature. 

Here, we present a study of electric and thermal conductivity in SrTi$_{1-x}$Nb$_x$O$_3$ at two different carrier concentrations ($n=3.1\times10^{20} cm^{-3}$ and  $n=1.8\times10^{20} cm^{-3}$). Despite the dominance of the lattice contribution to the heat transport in strontium titanate \cite{Lin2014b,martelli2018,Jiang2022}, we succeeded in extracting the electronic contribution to heat transport by exploiting the differentiating effect of the magnetic field on phonons and electrons. Such a method was employed previously in the case of bismuth and antimony \cite{white1958,Uher1974,Jaoui2021,jaoui2022}. We found that $WT$ follows Eq. \ref{T-square2} and $L_0B > A$. Thus T-square resistivity in SrTi$_{1-x}$Nb$_x$O$_3$ cannot be distinguished from other metals in which the e-e origin of the T-square resistivity is uncontested. This leads us to conclude that T-square (electric and thermal) resistivity can be caused without Umklapp as a consequence of the T-square decrease in the amplitude of the (momentum and energy) diffusivity in a Fermi liquid caused by fermion-fermion scattering. A comprehensive theory of this phenomenon is yet to be elaborated. 

\begin{figure*}[h!]
%\centering
\includegraphics[width=17cm]{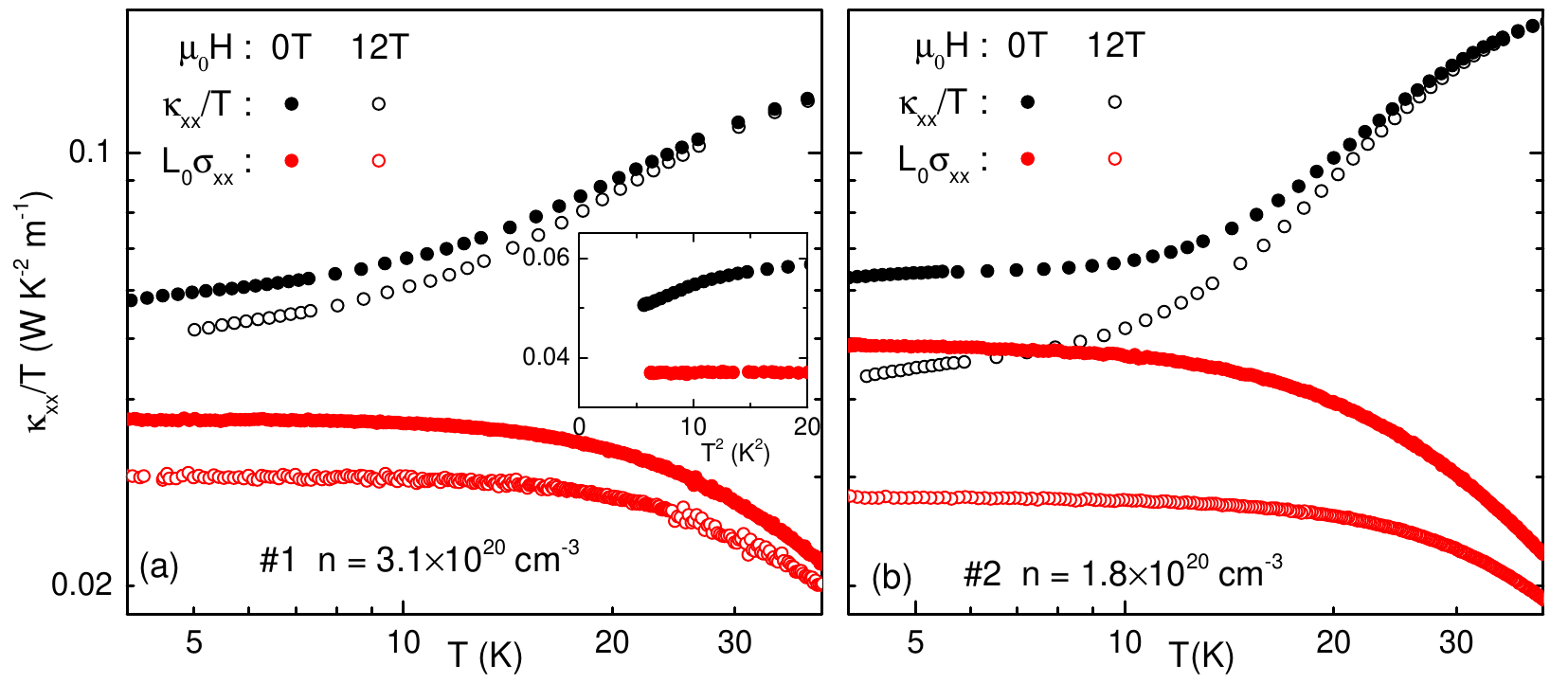}
\caption{\textbf{Thermal conductivity in Nb-doped SrTiO$_3$:} Thermal conductivity divided by temperature ($\kappa_{xx}/T$) at B= 0 and B= 12 T compared with the electrical conductivity multiplied by the Sommerfeld value ($L_0 \sigma_{xx}$) in sample \#1 (a) and in sample \#2 (b). $\kappa_{xx}/T$ increases with warming, because of the phonon contribution, which rises faster than $T$. $L_0 \sigma_{xx}$, which is a rough estimate of electronic contribution to $\kappa_{xx}/T$ decreases with warming due to the reduction of electrical conductivity by inelastic scattering. Note the reduction induced by magnetic field in both. The inset is a zoom on the low-temperature data, showing that they tend to join in the zero-temperature limit.}
\label{fig1}
\end{figure*}

Fig \ref{fig1} shows the temperature dependence of the total thermal conductivity in  two samples of SrTi$_{1-x}$Nb$_x$O$_3$. As discussed in the supplementary material \cite{SM}, the temperature dependence of electrical resistivity in our samples is comparable to what has been reported previously for this doping level and their residual resitivity is close to the lower end of the spectrum. In order to see the relative share of the electronic and the phononic contributions to the total heat transport, $\kappa_{xx}/T$ is compared with $L_0\sigma_{xx}$, which represents the upper boundary of electronic thermal conductivity according to the WF law. With decreasing temperature, $\kappa_{xx}/T$ approaches $L_0\sigma_{xx}$. As the temperature tends to zero (see inset), they tend to join each other. $\kappa_{xx}/T$ and $L_0\sigma_{xx}$ are both modified by the presence of a perpendicular 12 T magnetic field . To quantify longitudinal conductivity in presence of magnetic field, we measured both the electrical and the thermal Hall resistivities and inverted the resistivity tensor. 

The field-induced decrease in electrical conductivity $\sigma_{xx}$ (i.e. the magnetoresistance) was the subject matter of a previous study \cite{Collignon2021}, which found that both longitudinal and transverse conductivity follow the behavior expected in the semi-classical picture: 
\begin{equation}
\sigma_{xx}=\frac{n e \mu }{1+\mu^2 B^2}
  \label{sigma_xx}
\end{equation}
\begin{equation}
\sigma_{xy}=\frac{ne \mu}{1+\mu^2B^2} \mu B
  \label{sigma_xy}
\end{equation}

\begin{figure*}[h!]
\includegraphics[width=17cm]{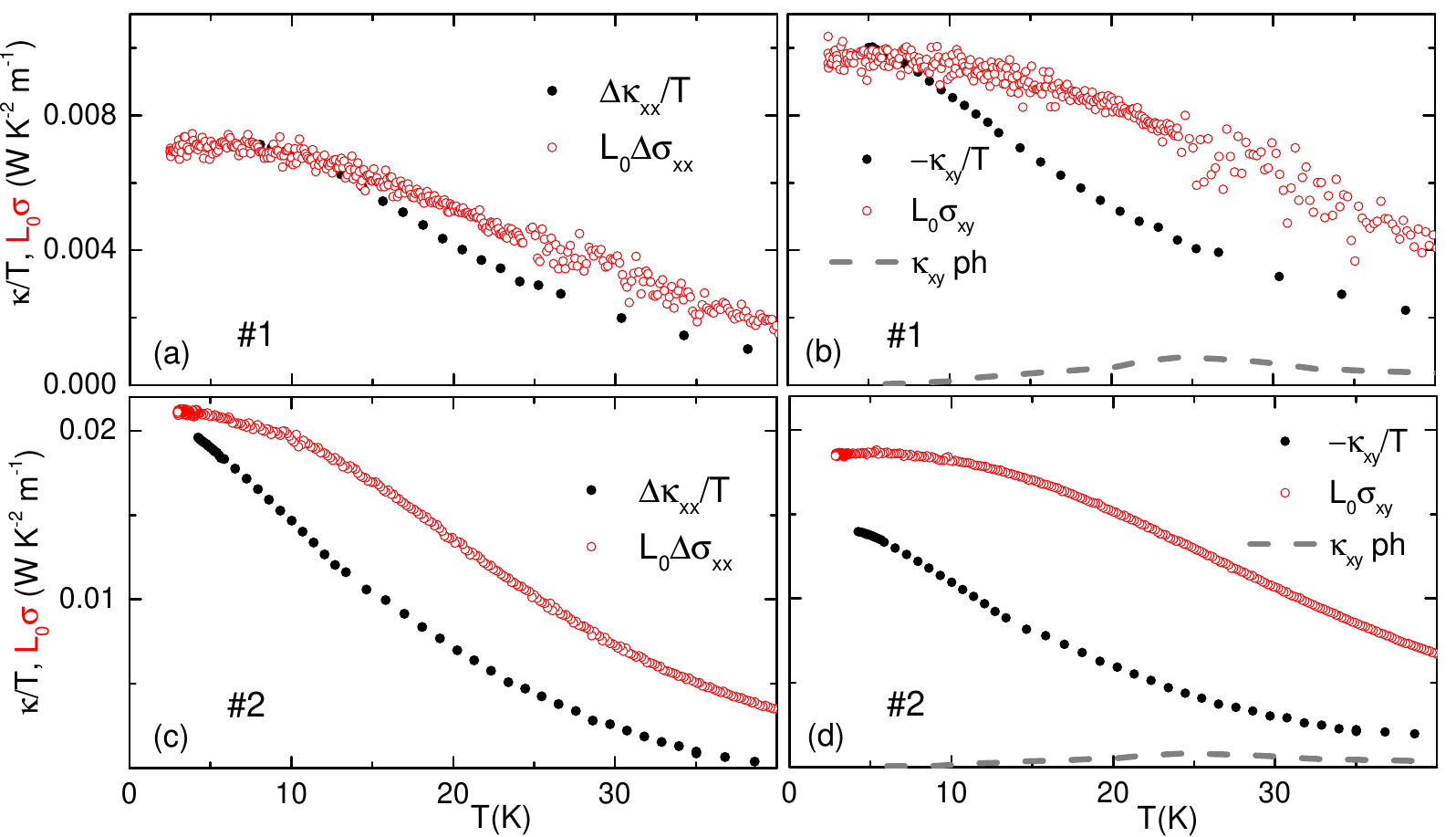}
\caption{\textbf{Longitudinal and Hall conductivity:} (a) The difference in longitudinal thermal conductivity divided by temperature between zero field and 12 T. ($\Delta \kappa_{xx} = \kappa_{xx}(0T)-\kappa_{xx}(12T)$) in sample \#1. Also shown is the difference in the longitudinal electric conductivity multiplied by the Sommerfeld value ($\Delta \sigma_{xx} = \sigma_{xx}(0T)-\sigma_{xx}(12T)$). (c)  Same for sample  \#2. (b) The transverse thermal conductivity $\kappa_{xy}$ divided by temperature, compared with the transverse electric conductivity $\sigma_{xy}$ multiplied by  $L_0$ (d) same for sample \#2. Also shown in (b) and (d) is  the $\kappa_{xy}/T$ caused by phonons in undoped pure STO \cite{XiaokangLi2020,Jiang2022}. }
\label{fig2}
\end{figure*}
Here, $n$ is the carrier density, and $e$ is the electron charge. Mobility, $\mu(B,T)$, is the only adjustable parameter depending on temperature and magnetic field. It monotonically decreases with increasing magnetic field and/or temperature. A remarkable (and poorly understood) fact about metallic STO is that the field dependence of mobility shows little dependence on the orientation of the magnetic field \cite{Collignon2021}. 

The thermal conductivity tensor $\overline{\kappa}$, on the other hand has an electronic $\overline{\kappa^e}$ and a lattice  $\overline{\kappa^{ph}}$ component in longitudinal:  $\kappa_{xx}(B) = \kappa^{e}_{xx}(B) + \kappa^{ph}_{xx}$. 

As in previous studies on semi-metals \cite{white1958,Uher1974,Jaoui2021}, one can separate the two components by assuming that the field dependence of the lattice thermal conductivity is negligible compared to the field dependence of electronic thermal conductivity. In insulating strontium titanate, thermal conductivity is purely phononic. There, at 24 K, a magnetic field of 12 T reduces $\kappa_{xx}$ at most by $7 \times 10^{-3}$ \cite{XiaokangLi2020} and generates a  finite thermal Hall conductivity of $\kappa^{ph}_{xy} \approx 0.09 W/K.m$. In our metallic samples, the effect of magnetic field on $\kappa^{e}_{xx}$ and the amplitude of $\kappa^{e}_{xy}$ (see below) are orders of magnitude larger.  

Fig \ref{fig2} shows the temperature dependence of the transverse thermal conductivity divided by temperature ($-\kappa_{xy}/T$). In the whole temperature range, it remains close (but smaller than $L_0\sigma_{xy}$), which is what is expected for the electronic part. The measured signal is much larger than $\kappa^{ph}_{xy}/T$ measured in insulating STO \cite{XiaokangLi2020} \footnote {In STO samples with a carrier density two orders of magnitude lower ($n \approx 10^{18} cm^{-3}$), $\kappa_{xy}/T$ is larger than $L_0\sigma_{xy}$, in contrast with the samples studied here, where the carrier density is two orders of magnitude larger ($n \approx 10^{20} cm^{-3}$ and the Hall angle is much smaller than unity. In this case, one does not expect to see a detectable phonon drag contribution to $\kappa_{xy}$ (See the supplement \cite{SM} for a more detailed discussion).}.

Thus, we can safely identify the field-induced change in thermal conductivity $\Delta \kappa_{xx}$ with the thermal magnetoresistance of electrons : 
\begin{equation}
  \Delta \kappa_{xx} = \kappa^e_{xx}(B=0)-\kappa^e_{xx}(B) 
  \label{deltakappa}
\end{equation}

Fig \ref{fig2} (a) and (c) compare  $\Delta \kappa_{xx}/T$ with $L_0\Delta\sigma_{xx}$. In both samples, these two quantities converge at low temperature and their difference grows with increasing temperature. This implies the validation of the WF law at zero temperature, a departure from it at finite temperature. The finite-temperature departure from the WF law is more significant in the sample with lower carrier density.

 \begin{figure*}
\centering
\includegraphics[width=18cm]{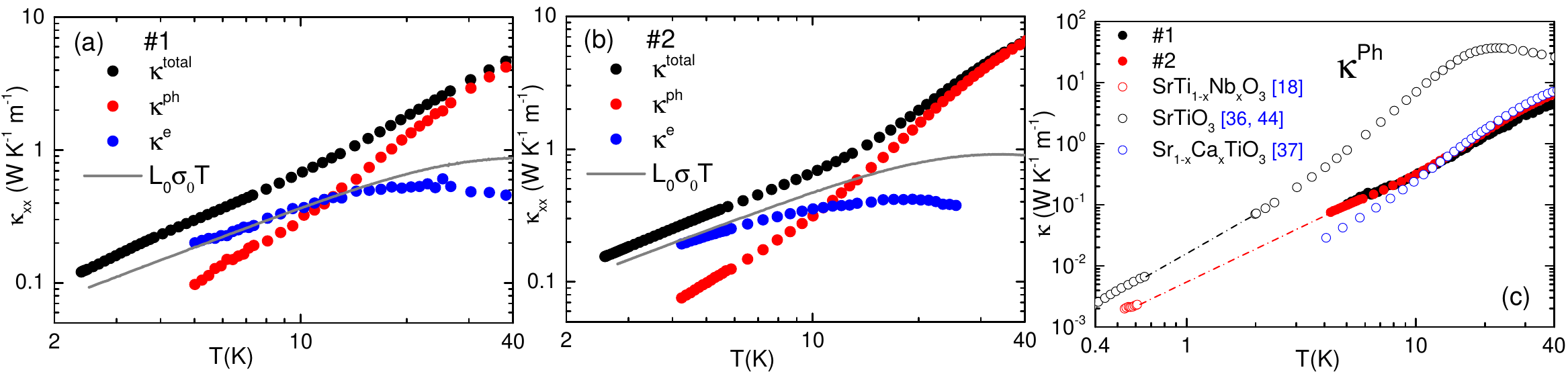}
\caption{\textbf{Electron and phonon contributions to the thermal conductivity} (a) The total thermal conductivity ($\kappa^{total}_{xx}$) and its  electronic ($\kappa^{e}_{xx}$) and phononic ($\kappa^{ph}_{xx}$) components as a function of temperature in sample \#1. Also shown is the electrical conductivity multiplied by L$_0$. (b) same for sample \#2. (c) Comparison of the phonon thermal conductivity in the two samples with total thermal conductivity of undoped STO \cite{martelli2018, rischau2021} with $\kappa ^{ph}_{xx}$ in SrTi$_{1-x}$Nb$_x$O$_3$ $n=2.6 \times 10^{20} cm^{-3}$ just above the superconducting transition \cite{Lin2014}  and with Sr$_{0.991}$Ca$_{0.009}$TiO$_3$ \cite{Jiang2022}. }
\label{fig3}
\end{figure*}

In order to extract the longitudinal electronic thermal conductivity at zero magnetic field ($\kappa^{e}_{xx} (B=0)$) from thermal magneto-conductivity ($\Delta \kappa_{xx}$), we need an additional assumption: At any given temperature, the field dependence of $\kappa^{e}_{xx}$, is similar to the field dependence of the electrical conductivity (expressed by Eq. \ref{sigma_xx}). Since the field-induced reduction in conductivity, in both thermal and electrical channels, is due to the same Lorentz force, this is a reasonable assumption. It implies that the Lorenz ratio ($L= \frac{\kappa^{e}_{xx}}{\sigma_{xx}T}$)  has a negligible field dependence. This assumption is consistent with our field-dependent data, which shows at a given temperature $\frac{L}{L_0}$ is less than unity, but its amplitude does not depend on magnetic field (See Fig. S2 in the supplement \cite{SM}). Thus, the magnetoresistance, which is not quadratic in magnetic field is set by the field dependence of residual resistivity and there is no detectable field-induced change in inetlastic scattering. This approach leads us to \cite{SM} :
\begin{equation}
  \kappa^{e}_{xx} (B=0) = \Delta\kappa_{xx} \frac{\sigma_{xx}(B=0)}{\Delta \sigma_{xx}}
  \label{kappa_e}
\end{equation}

Having quantified $\kappa^{e}_{xx}$ from our data, we can deduce $\kappa^{ph}_{xx}$ by subtracting the electronic component from the total conductivity. Fig \ref{fig3} shows the results. One can see in panels (a) and (b) that, above 20 $K$, $\kappa^{ph}_{xx}$ becomes an order of magnitude larger than $\kappa_{e}$. However, since $\kappa^{ph}_{xx}$ decreases faster than $\kappa^{e}_{xx}$ with cooling, the electron contribution becomes prominent below 10 $K$. It is almost equal to $L_0\sigma_0 T$ at low temperature, but becomes significantly lower at higher temperatures.

The extracted $\kappa^{ph}_{xx}$, shown in Fig \ref{fig3} (c), is significantly lower than $\kappa^{ph}_{xx}$ in undoped STO \cite{martelli2018}. As one can see in the figure, $\kappa^{ph}$ of our metallic samples, with about $1\%$ of Ti atoms replaced by Nb, is similar to the total $\kappa$ of insulating samples of Sr$_{1-x}$Ca$_{x}$TiO$_3$, with about $1\%$ of Sr atoms are replaced by Ca. In both cases, $\kappa^{ph}$ is reduced in comparison to pristine STO, because the substituting atoms are randomly distributed and their average distance is comparable with the order of magnitude of the wavelength of thermally excited phonons. The rough similarity between Nb doping (which brings mobile electrons) and Ca substitution (which does not), indicates that scattering by mobile electrons plays a minor role.
\begin{figure}[h!]
\centering
\includegraphics[width=8.5cm]{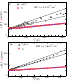}
\caption{\textbf{Electronic thermal resistivity:} (a) Electric and thermal resistivity as a function of the square of the temperature for sample $\#1$ (b) Same as in (a) for sample$\#2$. In both cases, $\rho$ and WT have the same intercept but different slopes. The two black solid lines show the lower and the upper limit to the slope of thermal resistivity in the two samples.}
\label{fig4}
\end{figure}

Let us now turn our attention to the electronic thermal resistivity, $WT$, obtained by inverting $\kappa^e_{xx}/T$.  Fig \ref{fig4} shows $\rho$ and $L_0 WT$ as a function of $T^2$ for the two samples. One can see that in both samples, Equations \ref{T-square1} and \ref{T-square2} hold.  $\rho_0$ and $L_0WT_0$ are identical at low temperature confirming the validity of the WF Law in the zero-temperature limit. In both samples, the slope of $L_0WT(T^2)$ ($B$ in equation \ref{T-square2}) is larger than the slope of $\rho(T^2)$ ($A$ in equation \ref{T-square2}).  This behavior is similar to what has been observed in semimetals (like W, WP$_2$, and Sb) and heavy-fermions (like UPt$_3$, and CeRhIn$_5$) (See Tab.\ref{table1}) and corresponds to what is theoretically expected in the e-e scattering picture \cite{Li-Maslov}.

The fermiology of doped strontium titanate has been the subject of several recent studies \cite{Marel2011,Lin2013,Allen2013,Lin2014,Fauque2022}. Experiments have confirmed that, as expected by band calculations \cite{Marel2011}, three bands associated with Ti orbitals are successively filled, as the doping increases. In the two samples studied here, the carrier density is such that the Fermi surface consists of three concentric pockets all three centered at the gamma point. The average radius of the outer pocket is bounded by the carrier density: $k_F^{max}<(3\pi^2 n )^{1/3}$. This yields 2.1 (1.7) nm$^{-1}$ in sample \#1 (\#2). The width of the Brillouin zone is $G=\frac{2\pi}{a}=16.1 nm^{-1}$, where $a=0.3905$ nm is the lattice parameter. Since $k_F^{max} < \frac{G}{4}$, Umklapp events cannot occur. This distinguishes metallic strontium titanate from other metals displaying T-square $\rho$ and $WT$ with amplitudes linked to each other by the WF law.

The other Fermi liquid with a T-square thermal resistivity in absence of Umklapp is normal liquid $^3$He \cite{Dobbs,Leggett_2016}. The dominant contribution to thermal conductivity (in the zero-temperature limit) is proportional to the inverse of temperature. This $\kappa T$ term is strictly equivalent to the inverse of $B$, the slope of $WT$ as a function $T^2$, and, as first calculated by Abrikosov and Khalatnikov \cite{Abrikosov_1959}, is proportional to the fermion-fermion scattering time, which quadratically decreases with temperature. Extracted from thermal conductivity, this scattering time was dubbed $\tau_{\kappa}$, and $\tau_{\kappa}T^2$ was extensively measured by Greywall \cite{greywall1984}. Theoretically, this quantity was computed by quantifying the Landau parameters of the Fermi liquid \cite{Dy1969,Sykes1970,wolfle1979}. The agreement between the theoretically-computed and the experimentally-measured $\kappa T$ and $\tau_{\kappa}T^2$  is within experimental uncertainty at saturating vapor pressure and less than a factor of 2 near the melting pressure.

Let us now see how metallic strontium titanate fits in this picture. $\tau_{\kappa}$ is given by \cite{greywall1984}:

\begin{equation}
\kappa= \frac{1}{3} \frac{C_V}{V_m} v_F^2 \tau_{\kappa}
  \label{tauk}
\end{equation}

Here, C$_V$ is the molar specific heat, V$_m$ is the molar volume and v$_F$  the Fermi velocity. This means that, in analogy with the case of normal liquid $^3$He \cite{greywall1984}, $B$, the prefactor of T-square thermal resistivity, is inversely proportional to $\tau_{\kappa}T^2$: 

\begin{equation}
\frac{1}{\tau_{\kappa}T^2}= \frac{v_F^2}{3} \frac{\gamma}{V_m} B
  \label{tauk2}
\end{equation}

Using the reported values of $\gamma$ (ranging from 1.55 to 1.9 mJ/mol.$K^2$ \cite{Lin2014b,McCalla,Ambler1966} at this doping level) and extracting the average Fermi wave-vector from carrier density, one can quantify the Fermi velocity and find  $\tau_{\kappa}T^2$. The results are listed in Tab. \ref{Table2}. Unsurprisingly, $\tau_{\kappa}T^2$ is orders of magnitude larger in STO than in $^3$He, which has a lower Fermi temperature and higher fermion-fermion collision cross section.

A more instructive basis for comparison is a dimensionless collision cross-section defined as \cite{wolfle1979,Pfitzner1985,Vollhardt1990,Behnia2022b}:   
 \begin{equation}
\zeta= \frac{\hbar E_F}{\tau_{\kappa}T^2k_B^2}
  \label{zeta}
\end{equation}

The amplitude of this quantity in a Fermi liquid is set by a combination of its Landau parameters \cite{Pfitzner1983,Pfitzner1985,Pftiznzer1987}. As seen in table \ref{Table2}, $\zeta$ in $^3$He, strongly correlated and close to both localisation and a magnetic instability \cite{Vollhardt1984},
%At saturating vapor pressure (melting pressure), the Landau parameters are $F^s_0=5.39(14.56)$, $F^s_1=9.30(88.47)$, $F^a_0=-0.695(-0.753)$ \cite{Vollhardt1990}. The negative $F^a_0$ and the large $F^s_1$ 
varies from 35 to 60. In contrast, $\zeta$ in STO, at this doping level (where the effective mass is close to four times the bare electron mass \cite{Lin2014b,collignon2019}), is $\approx 3$. 

\begin{table*}[ht!]
\begin{tabular}{|c|c|c|c|c|c|}
\hline
 Material &  $\rho_0\ (n\ohm \cdot cm)$  & $A\ (n\ohm \cdot cm \cdot K^{-2})$ & $B\ (n\ohm \cdot cm \cdot K^{-2}) $ & $B/A$ &  reference \\
\hline
W  & $(5.66 \pm 0.04) \times 10^{-2}$ & $(8.7 \pm 0.3) \times 10^{-4}$ \ &\ $(5.3 \pm 0.15) \times 10^{-3}$ \ &\ $ 6.1 \pm 0.4 $ &\cite{Wagner1971} \\
\hline
Sb  & 30 & 0.3 & 0.63 & 2.1 & \cite{Jaoui2021} \\
\hline
WP$_2$ &\ $4.7$ \  &\ $1.66 \times 10^{-2}$ \ &\ $7.56 \times 10^{-2}$ \ &\ $4.55$ &\cite{Jaoui2018} \\
\hline
\ CeRhIn$_5$ \ &\ $37$ \ &\ $21$ \ &\ $57$ \ &\ $2.7$ &\cite{Paglione2005} \\
\hline
\ CeCoIn$_5$ (B=7 T)\ &\ $370$ \ &\ $2600$ \ &\ $5500$ \ &\ $2.1$ &\cite{Paglione2006} \\
\hline
\ CeCoIn$_5$ (B=10 T)\ &\ $530$ \ &\ $900$ \ &\ $1900$ \ &\ $2.1$ &\cite{Paglione2006} \\
\hline
\ UPt$_3$  (a-axis)  & 230  & 590 & 905  &  1.5  &\cite{Lussier1994}\\
\hline
\ UPt$_3$  (c-axis)   &610  & 1600  & 2445 & 1.5&\cite{Lussier1994} \\
\hline
\ Nb:STO (S\#1)   &\ $6 \times 10^4$ \ &\ $33$ \ &\ $90 \pm 20 $ \ &\ $2.7 \pm 0.6$ & This work \\
\hline
\ Nb:STO (S\#2)   &\ $4.5 \times 10^4$ \ &\ $40$ \ &\ $185 \pm 20 $ \ &\ $4.6 \pm 0.5 $ &This work \\
\hline
\end{tabular}
\caption{\textbf{T-square resistivity in metals-} Residual resistivity $\rho_0$, and electrical (A) and thermal (B) T-square prefactors  and their ratio in several Fermi liquids. In the case of CeCoIn$_5$ \cite{Paglione2006}, the Fermi liquid behavior appears only in presence of a magnetic field larger than the upper critical field of the superconductor. $\frac{B}{A}$ is always found to be larger than unity, varying between 1.5 and 6.}
\label{table1}
\end{table*}

\begin{table*}[ht!]
\centering
\begin{tabular}{|c|c|c|c|c|c|c|}
\hline
System & $\kappa$ T (W.m) & v$_F$ (m/s)& k$_F$ (nm$^{-1}$) & E$_F$(K)& $\tau_{\kappa}T^2$ (ns.K$^2$) &  $\zeta $\\
\hline
$^3$He (saturating vapor pressure) & 2.9 $\times 10^{-4}$& 60 & 7.9 & 1.8 & 3.9 $\times 10^{-4}$ &  35 \\
\hline
$^3$He (melting pressure) & 7.3 $\times 10^{-5}$& 32.4 & 8.9 & 1.1& 1.4 $\times 10^{-4}$ &  58 \\
\hline
SrTi$_{3-x}$Nb$_x$O$_3$ (n=3.1$\times 10^{20}cm^{-3}$) &27 $\pm 6$  & 6 $\times 10^4$ & 2.1 &480 & 1.4 & 2.6 $\pm 0.6$ \\
\hline
SrTi$_{3-x}$Nb$_x$O$_3$ (n=1.8$\times 10^{20}cm^{-3}$) &13 $\pm 1.5$  & 5 $\times 10^4$ & 1.7 &330 & 0.8 & 3.2 $\pm 0.3$\\
\hline
\end{tabular}
\caption{\textbf{ Thermal and electronic properties of $^3$He and SrTi$_{3-x}$Nb$_x$O$_3$:} Thermal conductivity, fermion-fermion scattering and Fermi liquid properties in a strongly correlated and a weakly correlated fermionic system.}
\label{Table2}
\end{table*} 

Thus, not only the existence of the T-square thermal resistivity in metallic strontium titanate,  but also its amplitude can be accounted for by considering it as a Fermi liquid with moderate correlations. As for T-square electrical resistivity (at low temperatures, that is below the degeneracy temperature of electrons and the minimum energy of the soft TO phonons), it could be accounted for, assuming a rough recovery of the Wiedemann-Franz law in presence of disorder. However, the theory for such a scenario is yet to be elaborated. It may require including the gradient of momentum flow caused by disorder and phonon scattering. 

We thank L. Hechler, M. Feigel'man, X. Li, A. Marguerite, D. Maslov,  D. Vollhardt P. W\"olfle, and Z. Zhu for discussions. This work was supported by the Agence Nationale de la Recherche (ANR-19-CE30-0014-04), by Jeunes Equipes de l$'$Institut de Physique du Coll\`ege de France and by a grant attributed by the Ile de France regional council. S.J. acknowledges a grant from China Scholarship Council. 

\bibliography{ref}
\clearpage
\onecolumngrid
% Add 'S' to the numbering inside the supplement
\renewcommand{\thesection}{S\arabic{section}}
\renewcommand{\thetable}{S\arabic{table}}
\renewcommand{\thefigure}{S\arabic{figure}}
\renewcommand{\theequation}{S\arabic{equation}}
\setcounter{section}{0}
\setcounter{figure}{0}
\setcounter{table}{0}
\setcounter{equation}{0}
\begin{center}{\large\bf Supplemental Material for ``T-square electron thermal resistivity in metallic strontium titanate"}\\
\end{center}
\setcounter{figure}{0}
 \section{Materials and methods}
 \begin{figure}[h!]
\centering
\includegraphics[width=8.5cm]{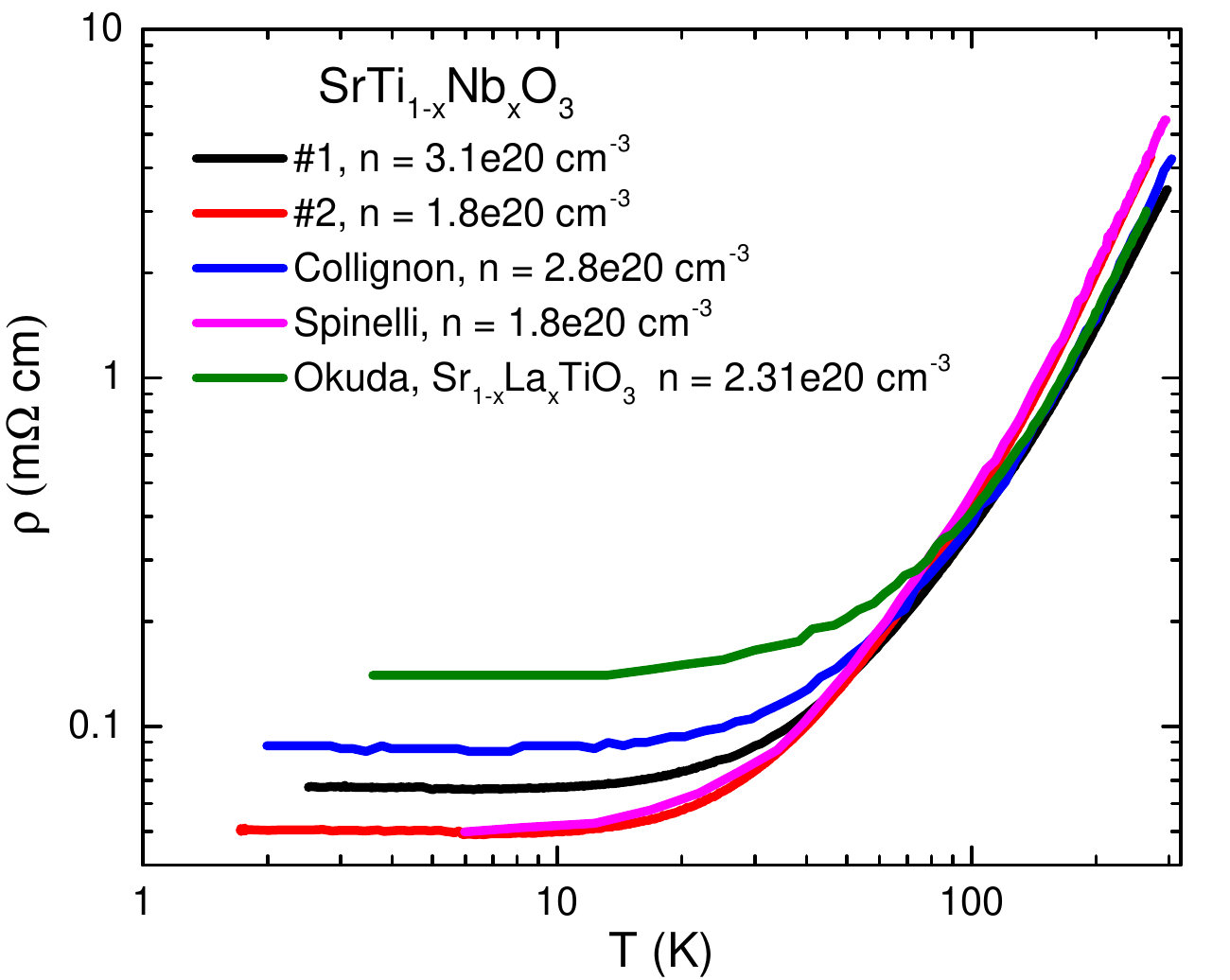}
\caption{\textbf{Electrical resistivity ($\rho$):} $\rho$ \it{vs.} temperature in log scale from 2 K to 300 $K$. Our samples ($\#1$ and $\#2$) are compared with those with similar carrier concentrations in previous studies of SrTi$_{1-x}$Nb$_x$O$_3$ \cite{Collignon2020,spinelli2010} and Sr$_{1-x}$La$_x$TiO$_3$ \cite{Okuda2001}. }
\label{figS1.1}
\end{figure}
SrTi$_{1-x}$Nb$_x$O$_3$ crystals were commercially provided by SurfaceNet GmbH. The nominal Nb content for sample $\#1$ ($\#2$) is $1\ wt\%$ ($0.5\ wt\%$). The expected carrier concentration for $\#1$ and $\#2$ is $3.3 \times 10^{20} cm^{-3}$ and $1.7 \times 10^{20} cm^{-3}$, respectively, in good agreement with the carrier concentration obtained by measuring the Hall coefficient ($\#1$ $n_H=3.1\times 10^{20} cm^{-3}$ and $\#2$ $n_H=1.8\times 10^{20} cm^{-3}$). Fig \ref{figS1.1} compares the temperature-dependence of their resistivity  with previous data on SrTi$_{1-x}$Nb$_x$O$_3$ \cite{Collignon2020, spinelli2010, Okuda2001} and Sr$_{1-x}$La$_x$TiO$_3$ \cite{Okuda2001} at similar carrier doping levels. Above 100 $K$, the resistivity of samples with the same carrier concentration is very similar. Below 80 $K$, they show different residual resistivities. Our samples tend to display a comparatively lower residual resitivity.

\begin{figure}[h!]
\centering
\includegraphics[width=8.5cm]{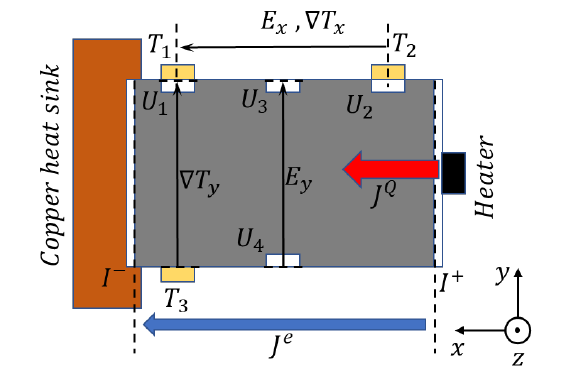}
\caption{\textbf{Sketch of the setup measurement :} T$_1$  T$_2$ and T$_3$ are three Cernox CX-1030 thermometers. The heater is a 1 $K\ohm$ chip resistor. All  connections were made using silver paste. 6 tin pads (white rectangles) are soldered on the sample for ohmic contacts. This setup allows us to obtain all transport coefficients.}
\label{figS1}
\end{figure}

Fig \ref{figS1} shows a sketch of the setup used to measure the electrical and thermal resistance reported in the main manuscript. Longitudinal ($\rho_{xx}=\frac{E_x}{J^e}$) and transverse resistivity ($\rho_{xy}=\frac{E_y}{J^e}$) are measured in absence of heat current ($J^{Q}$=0 and $J^e\ne$0). Longitudinal ($W_{xx}=\frac{-\nabla T_x}{J^Q}$) and transverse ($W_{xy}=\frac{-\nabla T_y}{J^Q}$) thermal resistivity are measured in absence of electrical current ($J^{e}$=0 and $J^Q\ne$0). In the measurement of $W_{xx}$ the temperature is the average temperature between $T_1$ and $T_2$. In the measurement of  $W_{xy}$, the average temperature is taken between $T_1$ and $T_3$. The  (electric or thermal) conductivity tensor is the inverse of the  (electric or thermal) resistivity tensor. Hence the longitudinal and transverse electric and thermal conductivity are respectively equal to $\sigma_{xx}=\frac{\rho_{xx}}{\rho_{xx}^2+\rho_{xy}^2}$ , $\sigma_{xy}=\frac{-\rho_{xy}}{\rho_{xx}^2+\rho_{xy}^2}$ and $\kappa_{xx}=\frac{W_{xx}}{W_{xx}^2+W_{xy}^2}$, $\kappa_{xy}=\frac{-W_{xy}}{W_{xx}^2+W_{xy}^2}$.

 \section{Field independence of the Lorenz ratio}
 \label{FieldLorenz}
Fig \ref{figS2} shows the field-dependent electric and thermal conductivity in sample $\# 1$. The  variation of thermal and electric conductivity caused by the magnetic field are similar. This is true for both the longitudinal and transverse response. The Lorentz force is thus the main source for the variation of thermal conductivity. The field-dependent Lorenz ratio in longitudinal and transverse are shown in Fig \ref{figS2} (c) and (f). The finite temperature departure of the Wiedemann Franz law  does not change with the magnetic field. This confirms our assumption of a field-independent Lorenz number. 

 \section{Absence of phonon drag thermal Hall effect}
 
Fig \ref{figS3} shows the Seebeck effect of sample $\#2$ compared with a SrTiO$_{3-x}$ sample with a carrier density two orders of magnitude lower $n = 1.6\times10^{18} cm^{-3}$. While a phonon drag peak is found in the low-doped material it is absent in sample $\#2$. Moreover samples $\#1$ and $\#2$ have a much smaller Hall angle compared with low doped SrTiO$_{3-\sigma}$ ($n = 1\times10^{18} cm^{-3}$). Therefore, no phonon drag contribution to  $\kappa_{xy}$ is expected  to arise in sample $\#2$ and also $\#1$ \cite{Jiang2022}.

 \section{Extracting electronic thermal conductivity from thermal and electrical magnetoresistance }
 
\begin{figure}[h!]
\centering
\includegraphics[width=17cm]{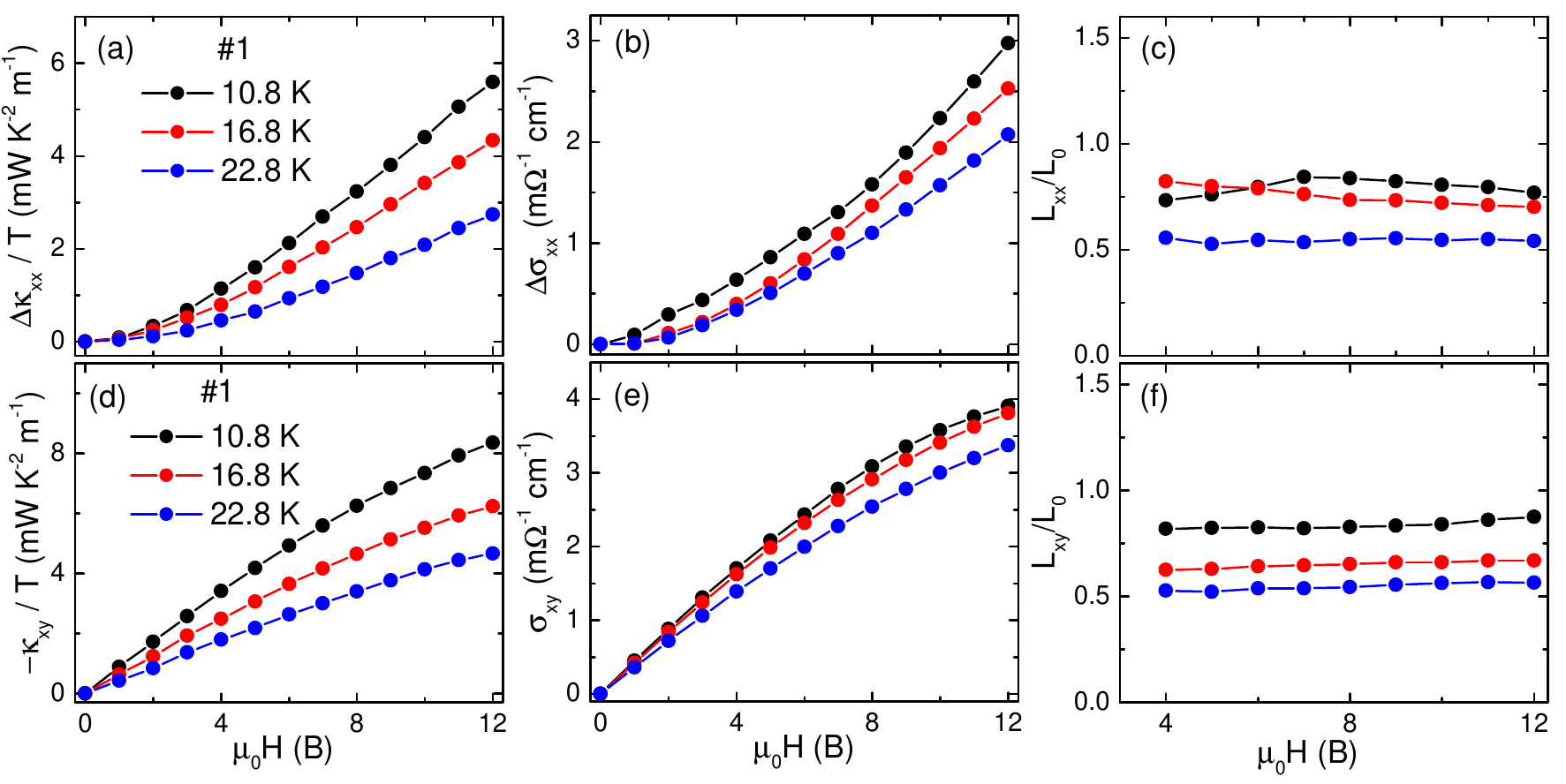}
\caption{\textbf{Field dependence of the electric and thermal conductivity:} (a)  Longitudinal thermal conductivity ($\Delta \kappa_{xx} = \kappa_{xx}(0)-\kappa_{xx}(B)$) divided by temperature, (b) Longitudinal electric conductivity ($\Delta \sigma_{xx} = \sigma_{xx}(0)-\sigma_{xx}(B)$) and (c) Longitudinal Lorenz ratio ($L_{xx}/L_{0} = \Delta \kappa_{xx}/T L_{0}\Delta \sigma_{xx}$) as a function of the magnetic field for three temperatures. (d), (e) and (f) are the same as (a), (b), and (c) in transverse configuration. The Lorenz ratios $L_{xx}$ and $L_{xy}$ are constant in field but smaller than L$_0$.}
\label{figS2}
\end{figure}

\begin{figure}[h!]
\centering
\includegraphics[width=17cm]{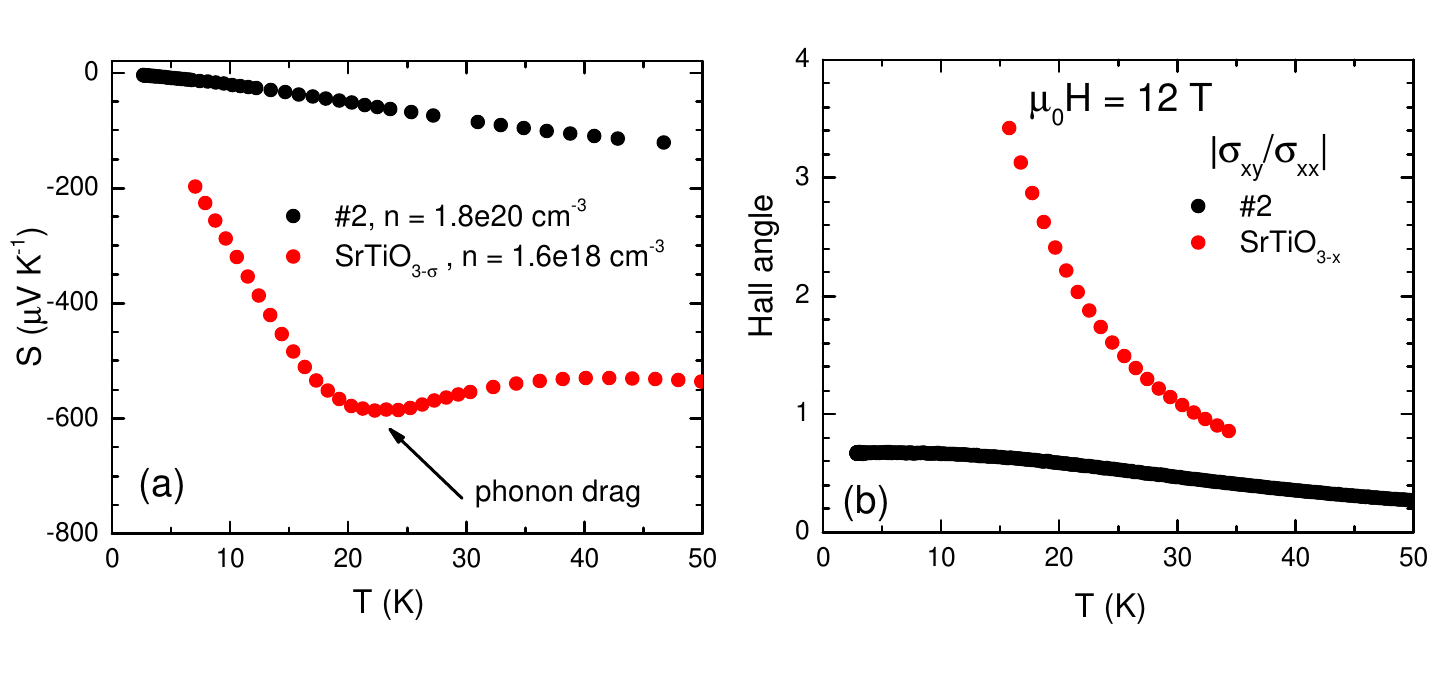}
\caption{\textbf{Seebeck effect and Hall angle:} (a) Seebeck coefficient ($S$), (b) Hall angle as a function of temperature for sample $\# 2$ ($n$=1.8e10$^{20}$cm$^{-3}$) and in a lightly doped SrTiO$_{3-\sigma}$ ($n$=1.6e10$^{18}$cm$^{-3}$). No phonon drag peak is observed in sample $\# 2$. The Hall angle in sample $\# 1$ is much lower than in lightly doped SrTiO$_{3-\sigma}$.}
\label{figS3}
\end{figure}

Two assumptions have been done to extract the temperature dependence of the electronic thermal conductivity. First we assume that the phonon contribution is independent of the magnetic field. The change in the total conductivity ($\kappa_{xx}$) is thus equal to the change in field of the electronic thermal conductivity : 
\begin{equation}
  \kappa_{xx}(B) = \kappa_{xx}^e(B)+\kappa_{xx}^{ph}
  \label{eqS1}
\end{equation}
%where $\kappa_{xx}^e(0) = L\sigma_{xx}(0)T$ and $\kappa_{xx}^e(B) = L\sigma_{xx}(B)T$. $L$ is the Lorenz number, which does not change with magnetic field and $T$ is absolutely temperature. Eventually, we can get the Lorenz number 
Second we assume that that the Lorenz number ($L$=$\frac{\kappa_{xx}^e}{\sigma_{xx}T}$) is independent of the magnetic field. This assumption is backed by our data (See Fig. \ref{figS2}c) in section \ref{FieldLorenz}). This implies that : 
\begin{equation}
  L = \frac{\Delta \kappa_{xx}}{\Delta \sigma_{xx}}\frac{1}{T}
  \label{eqS3}
\end{equation}
where $\Delta \kappa_{xx} = \kappa_{xx}(0)-\kappa_{xx}(B)$ and $\Delta \sigma_{xx} = \sigma_{xx}(0)-\sigma_{xx}(B)$. Using Eq \ref{eqS1} and \ref{eqS3}, we can extract the  electron component of the thermal conductivity at zero magnetic field :
\begin{equation}
  \kappa_{xx}^e(0) = \frac{\Delta \kappa_{xx}}{\Delta \sigma_{xx}}\cdot \sigma_{xx}(0)
  \label{eqS4}
\end{equation}
%\begin{equation}
 % \kappa_{xx}^{ph} =\kappa_{xx}(0) - \frac{\Delta \kappa_{xx}}{\Delta \sigma_{xx}}\cdot \sigma_{xx}(0)
 % \label{eqS5}
%\end{equation}

\end{document}